\documentclass[aip,reprint]{revtex4-1}

\usepackage{graphicx}
\usepackage{dcolumn}
\usepackage{bm}
\usepackage[utf8]{inputenc}
\usepackage[T1]{fontenc}
\usepackage{textcomp}
\usepackage{siunitx}
\usepackage{xcolor}

\begin{document}

\title{Nonlinear losses in magnon transport due to four-magnon scattering} 

\author{Tobias Hula}
\affiliation{Institut f\"{u}r Ionenstrahlphysik und Materialforschung, Helmholtz-Zentrum Dresden-Rossendorf, D-01328 Dresden, Germany}
\affiliation{Institut f\"{u}r Physik, Technische Universit\"{a}t Chemnitz, 09107 Chemnitz, Germany}

\author{Katrin Schultheiss} \email{k.schultheiss@hzdr.de}
\affiliation{Institut f\"{u}r Ionenstrahlphysik und Materialforschung, Helmholtz-Zentrum Dresden-Rossendorf, D-01328 Dresden, Germany}

\author{Aleksandr Buzdakov}
\affiliation{Institut f\"{u}r Ionenstrahlphysik und Materialforschung, Helmholtz-Zentrum Dresden-Rossendorf, D-01328 Dresden, Germany}

\author{Lukas K\"{o}rber}
\affiliation{Institut f\"{u}r Ionenstrahlphysik und Materialforschung, Helmholtz-Zentrum Dresden-Rossendorf, D-01328 Dresden, Germany}
\affiliation{Fakult\"{a}t Physik, Technische Universit\"{a}t Dresden, 01062 Dresden, Germany}

\author{Mauricio Bejarano}
\affiliation{Institut f\"{u}r Ionenstrahlphysik und Materialforschung, Helmholtz-Zentrum Dresden-Rossendorf, D-01328 Dresden, Germany}

\author{Luis Flacke}
\author{Lukas Liensberger}
\author{Mathias Weiler}
\affiliation{Walther-Meißner Institute, Bayerische Akademie der Wissenschaften, 85748 Garching, Germany}
\affiliation{Physik-Department, TU M\"{u}nchen, 80799 Munich, Germany}

\author{Justin M. Shaw}
\affiliation{Quantum Electromagnetics Division, National Institute of Standards and Technology, Boulder, Colorado 80305, USA}

\author{Hans T. Nembach}
\affiliation{Quantum Electromagnetics Division, National Institute of Standards and Technology, Boulder, Colorado 80305, USA}
\affiliation{JILA, University of Colorado, Boulder, Colorado 80309, USA}

\author{J\"{u}rgen Fassbender}
\author{Helmut Schultheiss}
\affiliation{Institut f\"{u}r Ionenstrahlphysik und Materialforschung, Helmholtz-Zentrum Dresden-Rossendorf, D-01328 Dresden, Germany}
\affiliation{Fakult\"{a}t Physik, Technische Universit\"{a}t Dresden, 01062 Dresden, Germany}

\date{\today}

\begin{abstract}

We report on the impact of nonlinear four-magnon scattering on magnon transport in microstructured Co$_{25}$Fe$_{75}$  waveguides with low magnetic damping. We determine the magnon propagation length with microfocused Brillouin light scattering over a broad range of excitation powers and detect a decrease of the attenuation length at high powers. This is consistent with the onset of nonlinear four-magnon scattering. Hence, it is critical to stay in the linear regime, when deriving damping parameters from the magnon propagation length. Otherwise, the intrinsic nonlinearity of magnetization dynamics may lead to a misinterpretation of magnon propagation lengths and, thus, to incorrect values of the magnetic damping of the system.
\end{abstract}


\maketitle 

In the growing field of magnonics,\cite{Neusser2009, Kruglyak2010, Lenk2011, Chumak2015} one aims at the use of magnons, the excitation quanta in  magnetically ordered systems, to transport and process information. In order to allow for coherent long-distance transport of signals in complex magnonic networks, the search for materials with low magnetic damping was reinitiated. In 2016, ferromagnetic resonance (FMR) measurements of continuous films revealed  intrinsic damping values as low as $(5\pm1.8)\times 10^{-4}$ for the conductor Co$_{25}$Fe$_{75}$,\cite{Schoen2016} approaching values found for the ferrimagnetic insulator yttrium iron garnet\cite{Sun2013}, with the added benefit of seminconductor compatibility. As a consequence, studies of the propagation characteristics in Co$_{25}$Fe$_{75}$ microstructures  followed, comparing damping values obtained from FMR to those derived from magnon propagation lengths.\cite{Koerner2017, Talalaevskiy2017, Flacke2019} In Ref.~\onlinecite{Koerner2017}, a 2.5 times higher damping is reported for magnon transport measurements than for FMR analysis of extended films, which is attributed to significant extrinsic contributions to the magnetic damping in the microstructured sample, such as local inhomogeneities and two-magnon scattering. 

However, one must consider that even in a perfect crystal, higher-order nonlinear scattering can lead to a significant increase of losses. The efficiency of such nonlinear processes depends on the population of the initial and the final magnon states. Hence, if fewer final states are available, {\it e.g.}, by patterning the continuous film to a microstructured waveguide, a higher population of  final states is achieved more easily, and the threshold for nonlinearities is reduced strongly.\cite{Schult12} This is even more the case in materials with low intrinsic damping. As a direct consequence, the efficiency of magnon transport certainly depends on the excitation amplitude, {\it i.e.}, whether the system is operated in the linear or nonlinear regime. One may think that in order to obtain most effective magnon transport, it is necessary to drive the excitation of magnons with the maximum power available. But counterintuitively, this may lead to increased losses due to nonlinear magnon scattering. 

In this Letter, we demonstrate that the application of high excitation powers above the threshold for nonlinear scattering may easily result in an underestimation of the magnon propagation lengths. Using spatially resolved Brillouin light scattering (BLS) microscopy of magnon transport in a Co$_{25}$Fe$_{75}$ waveguide, we  prove that magnon propagation lengths are significantly larger if excited in the linear compared to the nonlinear regime. 

\begin{figure}
\includegraphics[width=8.5cm]{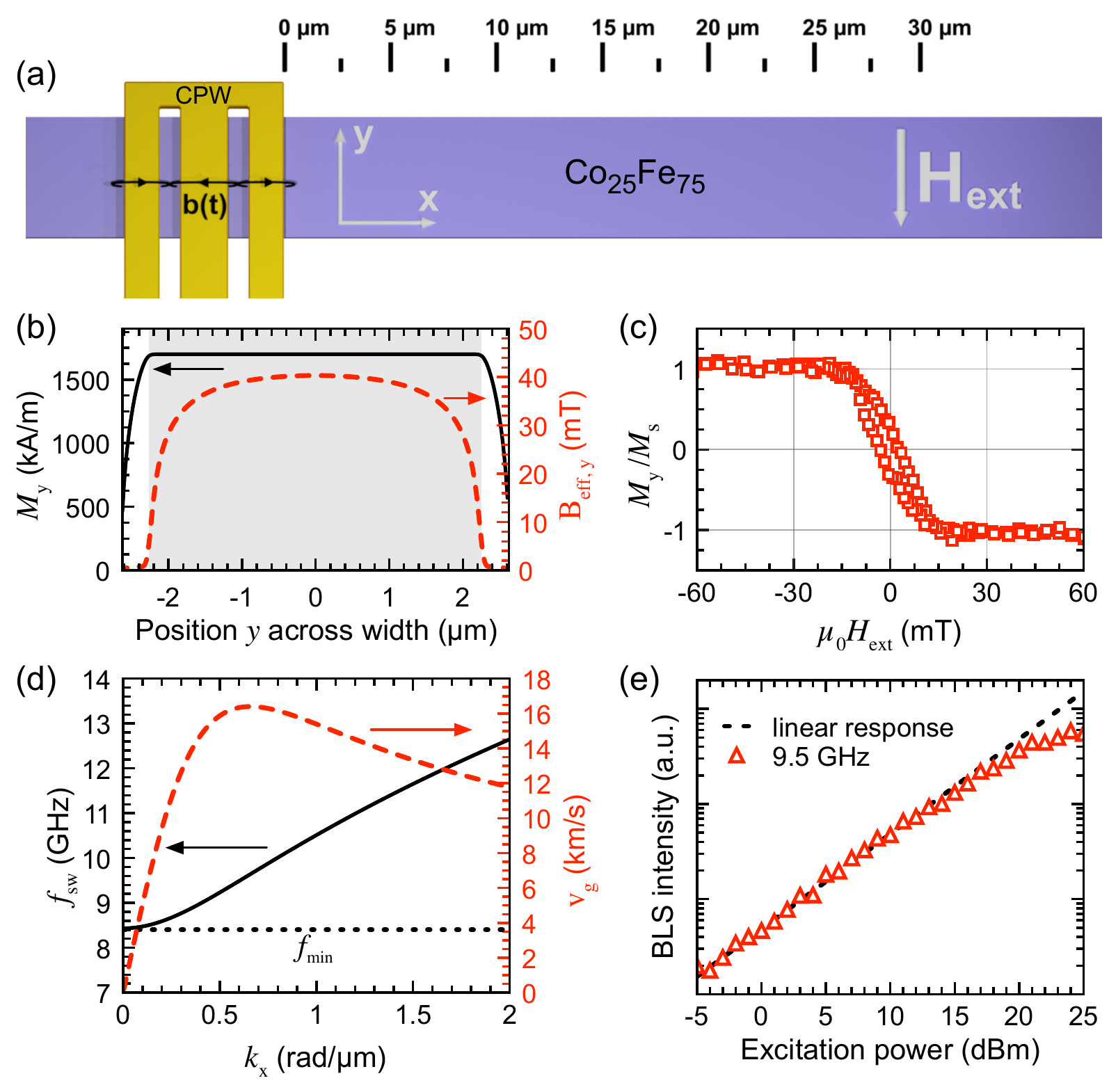}
\caption{\label{fig:sample} (a) Schematic of the investigated microstructure. Blue depicts the \SI{5.25}{\micro\meter} wide Co$_{25}$Fe$_{75}$ waveguide, yellow shows the coplanar waveguide (CPW). (b) Micromagnetic simulation of the $y$-component of the magnetization (black solid line)  and of the effective magnetic field $B_\mathrm{eff, y}$ (red dashed line), both extracted along the width of the waveguide for $\mu_0 H_\mathrm{ext}=\SI{46}{\milli\tesla}$. (c) Hysteresis loop of the waveguide in hard axis direction measured by laser scanning MOKE in the center of the waveguide. (d) Calculated dispersion relation (black solid line) and group velocity (red dashed line). (e) BLS intensity measured in a  \SI{600}{\mega\hertz} window around the excitation frequency $f_\mathrm{rf}=\SI{9.5}{\giga\hertz}$ as a function of the excitation power in \SI{1}{\micro\meter} distance to the CPW.}
\end{figure}

As illustrated in Fig.~\ref{fig:sample}a, we study a \SI{5.25}{\micro\meter} wide and \SI{60}{\micro\meter}  long magnon waveguide that was patterned from a Pt(3)/Cu(3)/Co$_{25}$Fe$_{75}$(26)/Cu(3)/Ta(3) multilayer (all thicknesses in nanometers)  using electron beam lithography, sputter deposition, and subsequent lift-off. In a previous study, we reported an intrinsic damping for this metallic thin film of $\alpha_{0}  \leq 3.18  \times 10^{4}$ in out-of-plane geometry. \cite{Flacke2019} In a second step, we patterned a Cr(5)/Au(70) film to a coplanar waveguide (CPW) with a  \SI{2.1}{\micro\meter} wide center conductor, separated by \SI{960}{\nano\meter} from the \SI{1.5}{\micro\meter} wide ground planes. The magnon waveguide extends up to \SI{38}{\micro\meter} on one side of the CPW. Microwave (rf) currents running through the CPW allow for the excitation of magnons with well-defined frequencies and high amplitudes, easily reaching the nonlinear regime. 

In order to achieve magnon propagation in Damon-Eshbach configuration,\cite{Damon} an external magnetic field of $\mu_0 H_\mathrm{ext}=\SI{46}{\milli\tesla}$ was applied perpendicularly to the long axis of the waveguide.  At this field, the magnetic moments align parallel to the external field over a \SI{4.5}{\micro\meter} wide center region, whereas the demagnetization field causes canting of the magnetization at the edges. This is confirmed by micromagnetic simulations using MuMax3\cite{mumax} (black solid line in Fig.~\ref{fig:sample}b). Furthermore, laser scanning magneto-optic Kerr effect measurements confirm that the magnetization in the center of the magnon waveguide is saturated for $\mu_0 H_\mathrm{ext}>\SI{20}{\milli\tesla}$ (Fig.~\ref{fig:sample}c). Hence, magnon propagation along the waveguide is governed by the characteristic dispersion with $\vec{k} \perp \vec{M}$ as plotted by the black solid line in Fig.~\ref{fig:sample}d, together with the corresponding group velocities $v_\mathrm{g} = 2\pi\frac{\delta f_\mathrm{sw}}{\delta k}$ (red dashed line). The calculation follows the formalism of Kalinikos and Slavin \cite{Kalinikos1990} assuming a quantization of the wave vector $k_\mathrm{y}$ across the stripe with an effective width of \SI{4.1}{\micro\meter} and an effective magnetic field of $\mu_0 H_\mathrm{eff}=\SI{40}{\milli\tesla}$, both obtained from micromagnetic simulations (Fig.~\ref{fig:sample}b). For our calculations and micromagnetic simulations, we assume material parameters as in Ref.~\onlinecite{Flacke2019}: $A_\mathrm{ex}=\SI{26}{\pico\joule/\meter}$, $g=2.051$, thickness $t=\SI{26}{\nano\meter}$, and $M_\mathrm{s}=\SI{1700}{\kilo\ampere/\meter}$. $M_\mathrm{s}$ is reduced compared to $\SI{1870}{\kilo\ampere/\meter}$ reported in Ref.~\onlinecite{Flacke2019} to account for heating due to the scanning BLS laser.\cite{Schult2008} 



In order to analyze the influence of nonlinear scattering on magnon propagation, we employed  Brillouin light scattering microscopy ({\textmu}BLS).\cite{Sebastian, Wagner16} This technique relies on the inelastic scattering of light and magnons followed by spectral analysis of the  scattered light via a high-resolution interferometer. The incident light, focused with a microscope objective onto the sample, yields a spatial resolution of about \SI{350}{\nano\meter}. The advantage of this technique compared to, {\it e.g.}, time-resolved magneto-optical Kerr effect or ferromagnetic resonance, lies in its capability to detect not only coherent but also incoherent magnons. This means we do not only have access to magnons that are directly excited by microwave fields but also to magnons created by nonlinear processes.

Such nonlinear interactions occur above a critical magnon amplitude. In our experiments, we determine this threshold in terms of a critical microwave power directly supplied by a signal generator without using any amplifiers. To this end, we apply a fixed microwave frequency of $f_\mathrm{rf}=\SI{9.5}{\giga\hertz}$ to the CPW and gradually increase the microwave power from
\SI{-5}{\deci\bel\of{m}} to $\SI{25}{\deci\bel\of{m}}$. We measure the magnon intensity integrated in a \SI{600}{\mega\hertz} window around the excitation frequency in \SI{1}{\micro\meter} distance from the antenna (red triangles in Fig.~\ref{fig:sample}e).

In the range between \SI{-5}{\deci\bel\of{m}} and $\SI{15}{\deci\bel\of{m}}$, the response of the magnetic system linearly follows the increasing excitation power (dashed black line). Above \SI{15}{\deci\bel\of{m}}, however, the measured intensities deviate from the extrapolated linear response. In this regime, the intensity is lower than would be expected from the increasing input power, indicating the onset of nonlinear scattering processes, which we will elucidate in more detail below.

Nonlinear processes have extensively been studied in a variety of publications, both experimentally and theoretically.\cite{Suhl1957, Demidov2009, HamiltonForm, Demidov2011, Schult12, Bauer2014, Schultheiss2019} In general, several nonlinear multi-magnon scattering processes, often referred to as \textit{Suhl instabilities}\cite{Suhl1957}, can cause a reduction of the population of directly excited magnons, all of which naturally obey energy and momentum conservation. Considering the dispersion relation in Fig.~\ref{fig:sample}d, the lowest order process of three-magnon scattering (one magnon with $f_\mathrm{rf}$ splits in two magnons with $f_\mathrm{1}+f_\mathrm{2}=f_\mathrm{rf}$) can be excluded in our system  since no states exist to satisfy this condition for $f_\mathrm{rf}=\SI{9.5}{\giga\hertz}$.

\begin{figure}
\includegraphics[width=8.5cm]{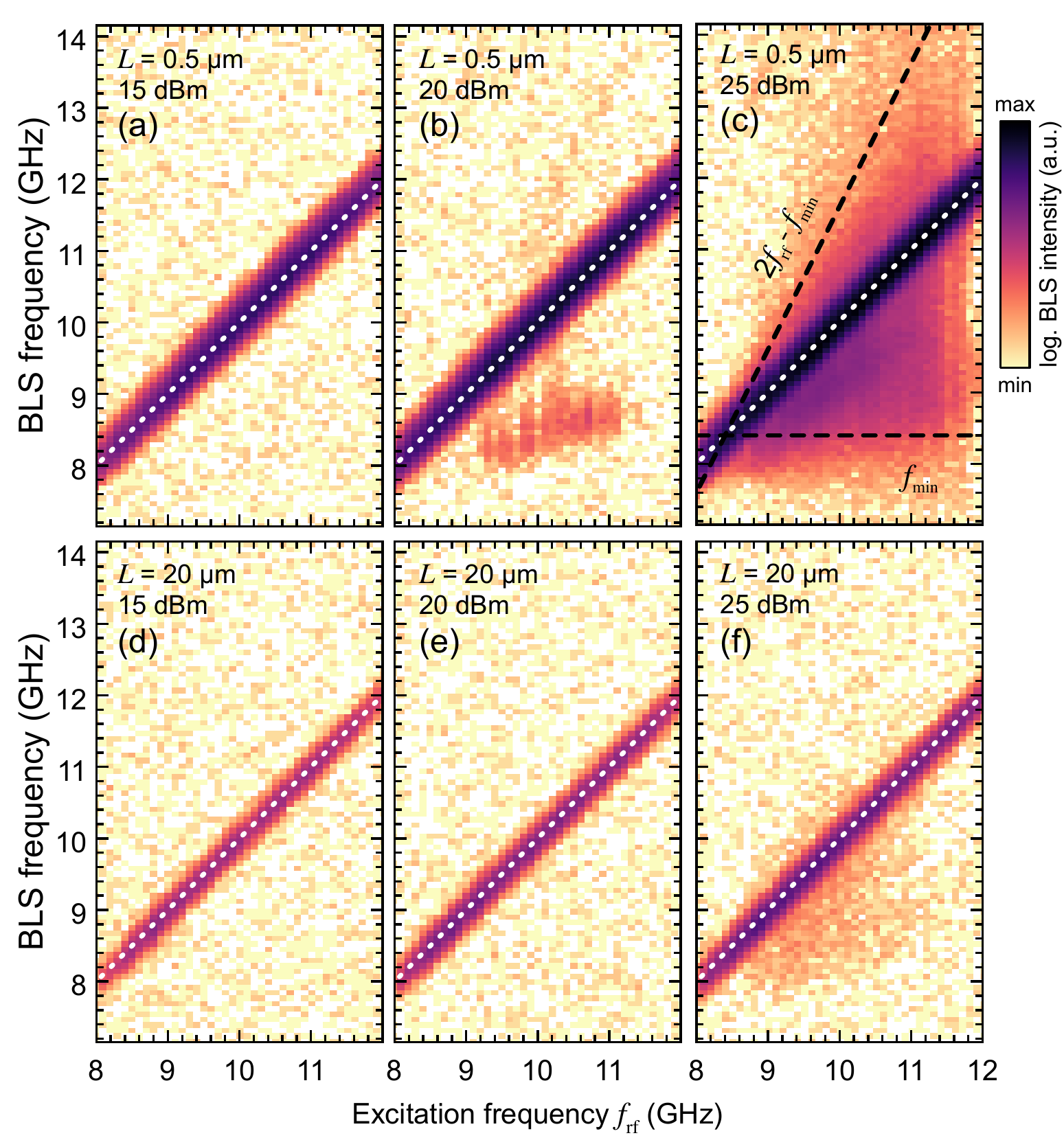}
\caption{\label{fig:RFmaps} {\textmu}BLS spectra measured in \SI{0.5}{\micro\meter} (a-c) and \SI{20}{\micro\meter} (d-f) distance to the antenna and at \SI{15}{\deci\bel\of{m}} (a,d), \SI{20}{\deci\bel\of{m}} (b,e), and \SI{25}{\deci\bel\of{m}} (c,f) microwave power. At low powers, the linear response of the magnetization dynamics to the external input dominates (white dotted lines). Especially for the highest power and close to the antenna, four-magnon scattering is apparent, which causes a significant broadening of the excitation spectra. All BLS intensities are plotted on the same logarithmic color scale.}
\end{figure}

The next higher order nonlinear process is given by four-magnon scattering: two primary magnons, that are excited at $f_\mathrm{rf}$, scatter and form two secondary magnons with $f_\mathrm{rf}\pm\delta f$, which is easily possible in our system due to the quasi-linear magnon dispersion over a wide range of wave vectors. To demonstrate that the intensity reduction in Fig.~\ref{fig:sample}e  is indeed in accordance with four-magnon scattering, we measure BLS spectra at three different microwave powers (\SI{15}{\deci\bel\of{m}}, \SI{20}{\deci\bel\of{m}}, and \SI{25}{\deci\bel\of{m}}) and in two distances $L$ from the antenna (\SI{0.5}{\micro\meter} and \SI{20}{\micro\meter}). The results are summarized in Fig.~\ref{fig:RFmaps}. Each column in the color maps shows a BLS spectrum that was recorded for microwave frequencies ranging from \SI{8}{\giga\hertz} to \SI{12}{\giga\hertz}. The measured intensities are color coded on the same logarithmic scale. 

Close to the antenna, at \SI{15}{\deci\bel\of{m}} (Fig.~\ref{fig:RFmaps}a), the measured magnon frequencies ($y$-axis) linearly follow the excitation frequency $f_\mathrm{rf}$ ($x$-axis), as indicated by the white dotted line. For increasing powers (Fig.~\ref{fig:RFmaps}b,c), however, we see a significant broadening around $f_\mathrm{rf}$. The limits of this broadening are indicated by black dashed lines and match the limits set by the bottom of the magnon manifold ($f_\mathrm{min}$) and energy conservation ($2f_\mathrm{rf}-f_\mathrm{min}$). This broadening is in agreement with previous studies driven at high excitation powers close to the exciting antenna and is a well known evidence for four-magnon scattering.\cite{Schult12, Bauer2014} 

Note that the intensities of nonlinearly excited magnons above  $f_\mathrm{rf}$ (white dashed line in Fig.~\ref{fig:RFmaps}b,c) are lower than in the range from $f_\mathrm{min}$ to $f_\mathrm{rf}$. This can be related to the decreasing detection sensitivity of the BLS microscope with increasing wave vector, {\it i.e.}, with increasing frequency \cite{Sebastian}. Additionally, in first order approximation, the lifetime of magnons scales inversely with their frequency which reduces the overall population of states with higher frequencies\cite{Stancil} and thus leads to a smaller signal.  

As a next step, we study the influence of four-magnon scattering on the propagation characteristics in the system. At \SI{20}{\micro\meter} distance to the antenna, the detected magnon intensities are attributed only to propagation, which was confirmed by phase-resolved BLS measurements in a previous investigation.\cite{Flacke2019} Nonetheless, Fig.~\ref{fig:RFmaps}f reveals that four-magnon scattering can still be evident at the highest excitation power, revealing the significant impact of four-magnon scattering on  magnon propagation at high power levels. 

\begin{figure}
\includegraphics[width=8.5cm]{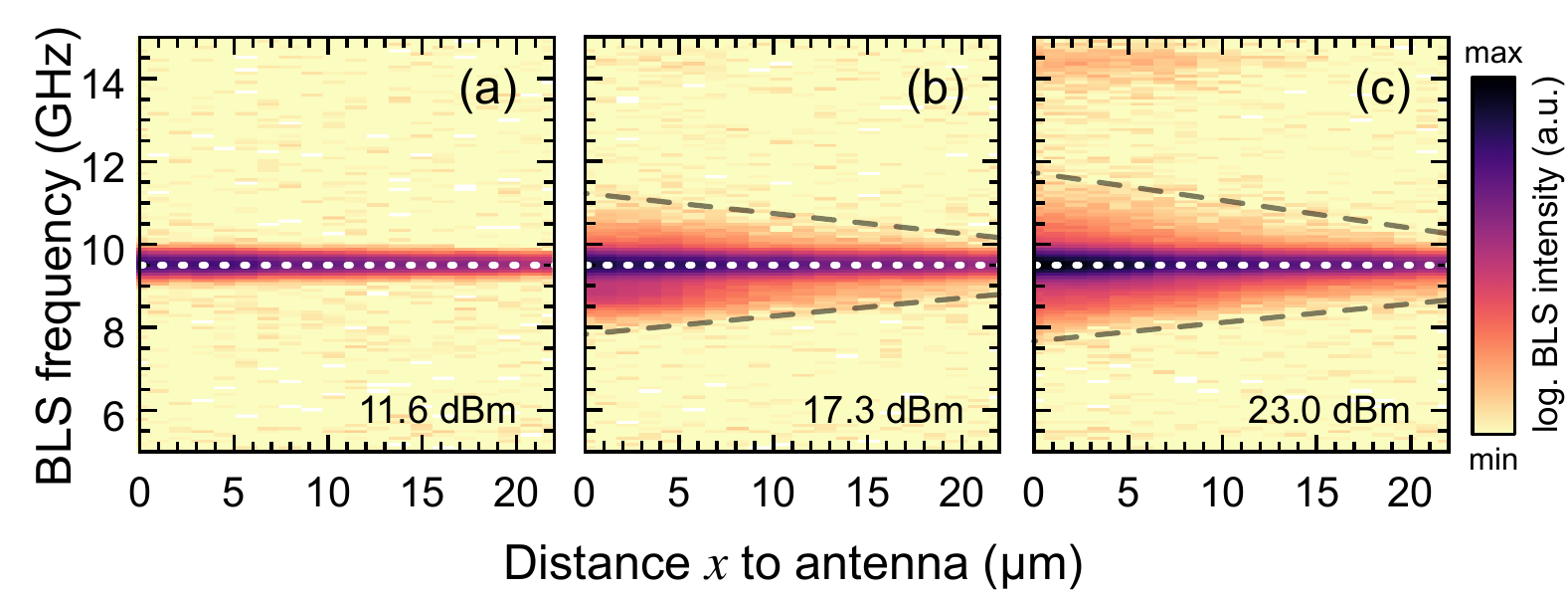}
\caption{\label{fig:spectraoverspace} {\textmu}BLS spectra measured along the waveguide for different excitation powers at a fixed excitation frequency $f_\mathrm{rf}=\SI{9.5}{\giga\hertz}$ (white dotted line). The grey dashed lines mark the distribution of magnons that are parametrically excited by four-magnon scattering.}
\end{figure}

To further elaborate on this phenomenon, we fix the excitation frequency at $f_\mathrm{rf}=\SI{9.5}{\giga\hertz}$ and measure magnon intensities as a function of distance $x$ to the antenna for three different excitation powers (Fig.~\ref{fig:spectraoverspace}). At \SI{11.6}{\deci\bel\of{m}}, we detect only magnons that are directly excited at \SI{9.5}{\giga\hertz} (white dotted line) and continuosuly decay along the waveguide. With increasing  power (Fig.~\ref{fig:spectraoverspace}b,c), the detected magnon spectrum broadens again, which is most pronounced close to the antenna and gets more and more narrow with increasing distance (indicated by dashed lines in Fig.~\ref{fig:spectraoverspace}b,c). One can assume that this spatial profile is defined by several contributions: First, primary magnons excited at $f_\mathrm{rf}$ by the antenna propagate along the waveguide with their amplitude decaying according to their damping. The same is true for secondary magnons that are  excited by nonlinear scattering above the nonlinear threshold. In the same manner, they propagate along the waveguide and suffer from damping. Second, four-magnon scattering causes energy flux from the primary to the secondary magnons. This energy flux provides an additional damping channel for the directly excited magnons at $f_\mathrm{rf}$ but also compensates losses of the secondary magnons. The coupling is based on dipole-dipole interaction and inversely proportional to the wave vector mismatch of the contributing states\cite{HamiltonForm}. Therefore, scattering into states $i$ with small $\delta k_\mathrm{i} = |k(\omega_\mathrm{rf})-k(\omega_\mathrm{i})|$ is expected to have lower threshold amplitudes, and to be more efficient. In the case of high excitation powers, this translates into the presence of these scattered states at larger distances from the antenna.

\begin{figure}
\includegraphics[width=8.5cm]{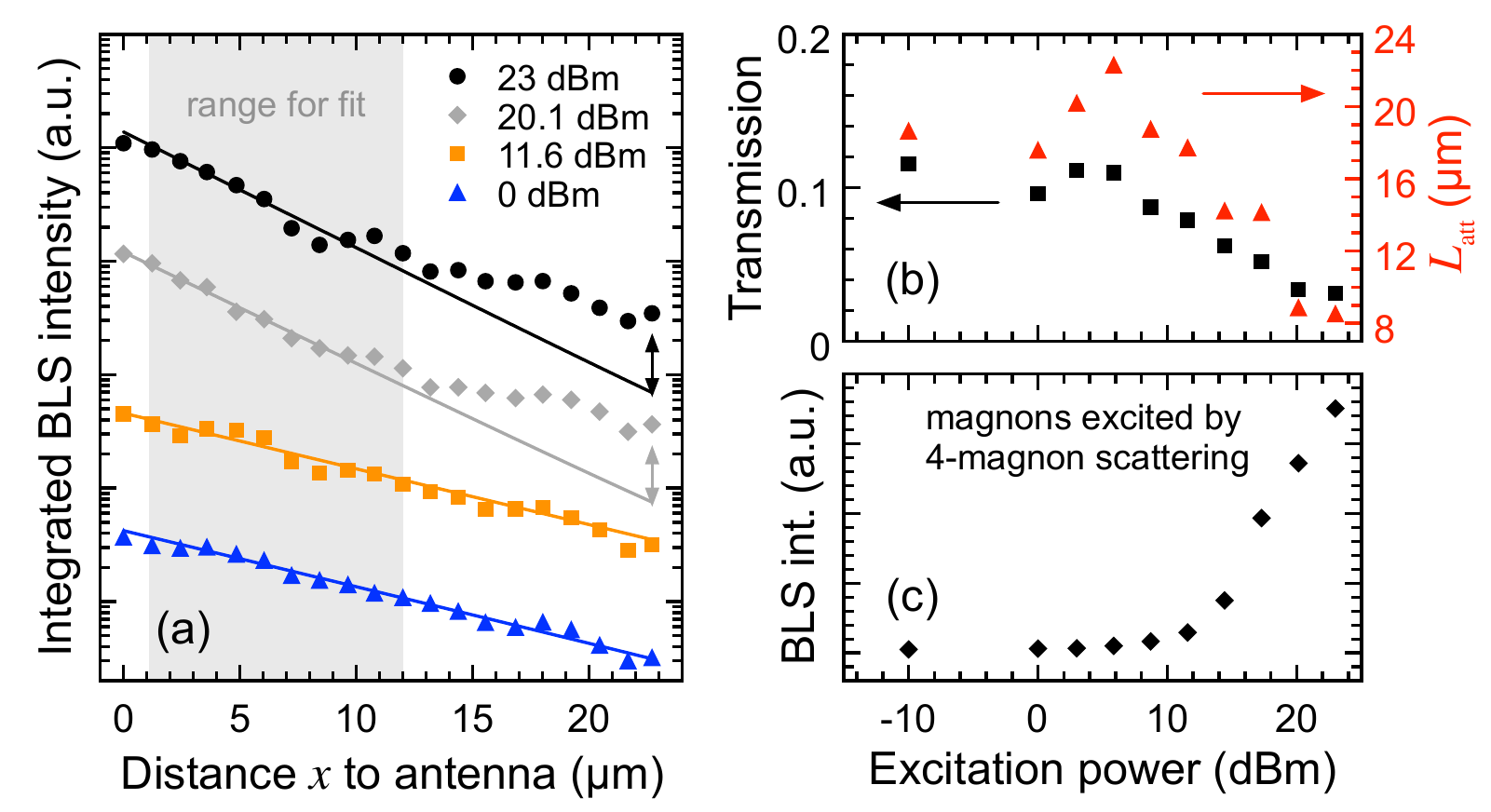}
\caption{\label{fig:decays} a) Integrated BLS intensities of magnons directly excited at $f_\mathrm{rf}=\SI{9.5}{\giga\hertz}$ as a function of the distance $x$ to the antenna for various RF powers. Solid lines show the results of exponential fits in the range between $x=\SI{1}{\micro\meter}$ and $x=\SI{12}{\micro\meter}$. Measurements at different powers are shifted vertically for clarity. b) Attenuation lengths (red triangles)  that were determined for different powers from the exponential fits in a. Relative transmission through the waveguide (black squares) given by the ratio of the integrated intensities measured in \SI{1}{\micro\meter} and \SI{22.7}{\micro\meter} distance to the antenna. c) BLS intensities only of magnons excited by four-magnon scattering (integrated in the ranges $f_\mathrm{BLS}<\SI{9.2}{\giga\hertz}$ and $f_\mathrm{BLS}>\SI{9.8}{\giga\hertz}$) and integrated over all measurement positions. }
\end{figure}

For a more detailed analysis of the nonlinear propagation characteristics, we repeat the same spatial measurements as in Fig.~\ref{fig:spectraoverspace} in a broader range of excitation powers. We integrate the BLS intensities over the entire frequency range of primary and secondary magnons and plot this as a function of the distance $x$ to the antenna. Figure~\ref{fig:decays}a shows the integrated intensities exemplarily for 0, 11.6, 20.1, and \SI{23}{\deci\bel\of{m}}. The data up to $x=\SI{12}{\micro\meter}$ suggest magnon propagation following a simple exponential decay of the intensity $I(x) = \exp(-2x/L_\mathrm{att})+c$ with $L_\mathrm{att}$ giving the attenuation length. The constant offset $c$ includes the thermal magnon background as well as thermal noise of the used photon counter. The factor two in the exponent accounts for the measurement of magnon intensities via BLS, which are proportional to the square of the magnon amplitudes. For fitting the data, we consider only the range  $\SI{1}{\micro\meter} \leq x \leq\SI{12}{\micro\meter}$ and plot the results as solid lines in Fig.~\ref{fig:decays}a.

At \SI{0}{\deci\bel\of{m}} and \SI{11.6}{\deci\bel\of{m}}, the fits agree well with the measured data for the entire range of distances from the antenna. However, at \SI{20.1}{\deci\bel\of{m}} and \SI{23}{\deci\bel\of{m}}, the fits strongly deviate from the experimental results for $x>\SI{12}{\micro\meter}$ and the measured intensities exceed the predictions of the exponential fits. This already indicates that nonlinear processes have a crucial impact on the propagation of magnons when driven above the critical threshold. Fitting data in a limited range close to the antenna at high excitation powers may lead to an underestimation of the actual decay lengths.

Red triangles in Fig.~\ref{fig:decays}b summarize the attenuation lengths $L_\mathrm{att}$ that we obtained from the exponential fits in the range $\SI{1}{\micro\meter} \leq x \leq\SI{12}{\micro\meter}$ for all excitation powers. From \SI{-10}{\deci\bel\of{m}} up to \SI{10}{\deci\bel\of{m}}, the attenuation lengths remain on a constant level around $\SI{19}{\micro\meter}$. For higher excitation powers, a pronounced reduction of the attenuation lengths occurs, reaching a minimum of \SI{8.5}{\micro\meter} at \SI{23}{\deci\bel\of{m}}. However, if we fit the BLS intensities for further distances from the antenna, {\it i.e.}, $x > \SI{12}{\micro\meter}$, all fits over the entire power range yield  propagation lengths in the range of $\SI{20}{\micro\meter}$.



As a different means to describe the total signal losses over the propagation length of \SI{22.7}{\micro\meter}, we calculated a transmission factor $T=I_f/I_i$, where $I_i$ ($I_f$) is the measured intensity at $x=\SI{1}{\micro\meter}$ ($x=\SI{22.7}{\micro\meter}$), respectively (black squares in Fig.~\ref{fig:decays}b). The power dependence of this parameter is consistent with the determined attenuation lengths. In the investigated range of microwave powers, the transmission is reduced by almost one order of magnitude, showing the significance of four-magnon scattering on the effective losses. 

To demonstrate the influence of four-magnon scattering more directly, we  integrated only the contributions of the secondary magnons, {\it i.e.}, we integrated the BLS signal in the ranges $f_\mathrm{BLS}<\SI{9.2}{\giga\hertz}$ and $f_\mathrm{BLS}>\SI{9.8}{\giga\hertz}$. Figure~\ref{fig:decays}c displays the intensity of those magnons, that are excited via four-magnon scattering, as a function of the excitation power. At lower powers, this signal stays within the noise level of the measurement. Only above \SI{12}{\deci\bel\of{m}}, contributions from four-magnon scattering drastically increase, which is consistent with the decrease of the measured attenuation lengths and transmission in Fig.~\ref{fig:decays}b and the measurement of the threshold for nonlinear four-magnon scattering in Fig.~\ref{fig:sample}e. 

In conclusion, we quantitatively studied the impact of nonlinear four-magnon scattering on long range magnon transport in a Co$_{25}$Fe$_{75}$ waveguide. 
The increase of propagation losses coincides with the onset of nonlinear four-magnon scattering. Measurements of the full magnon spectrum along the waveguide showed the presence of nonlinearities for propagation lengths exceeding $\SI{10}{\micro\meter}$. Our studies show that it is crucial to stay in the linear regime in order to quantify propagation lengths in systems with reduced dimensionality. If one is not aware that the excitation powers already reach the nonlinear regime, one may underestimate the performance of their structure, especially when comparing the results to ferromagnetic resonance measurements on continuous films. 
\newline

Financial support within DFG programme SCHU 2922/1-1 is acknowledged. K.S. acknowledges funding within the Helmholtz Postdoc Programme.  LL, LF and MW acknowledge funding by the DFG via projects WE5386/4-1 and WE5386/5-1. The samples were partially fabricated at the Nanofabrication Facilities (NanoFaRo) at the Institute of Ion Beam Research and Materials Research at HZDR. We thank B. Scheumann for the deposition of the Cr/Au film.

\section*{Data Availability}
The data that support the findings of this study are available from the corresponding author upon reasonable request.

\end{document}